\documentclass[11pt,a4paper]{article}

\usepackage[margin=1in]{geometry}

\usepackage{authblk}

\usepackage{cite}
\usepackage[cmex10]{amsmath}
\usepackage{amsthm}

\usepackage{subcaption}

\usepackage{amssymb,amsfonts}
\usepackage{algorithmic}
\usepackage{textcomp}
\usepackage{xcolor}
\usepackage{hyperref}
\usepackage{tikz}
\usetikzlibrary{shapes.geometric, arrows.meta, positioning, fit, backgrounds}

\begin{document}

\title{Meshless Domain Randomization via Explicit Parameter Perturbation of 3D Gaussian Splatting}

\author{Felipe Nunes Carbone de Carvalho}
\author{Joyce de Morais Souza}
\author{Alan de Aguiar}
\author{Charles Morphy D. Santos}
\author{João Paulo Gois}

\affil{\small \textit{Federal University of ABC (UFABC), Santo André, Brazil} \\
\texttt{felipe.nunes@aluno.ufabc.edu.br}, \texttt{\{joyce.morais, alan.aguiar, charles.santos, joao.gois\}@ufabc.edu.br}}

\date{}

\maketitle

\begin{abstract}
Domain Randomization (DR) is a standard technique for closing the Sim-to-Real gap, yet traditional DR pipelines rely on classical computer graphics rendering driven by polygon meshes. For complex organic subjects, such as insect specimens, extracting and rendering textured meshes is challenging. To address this issue, we propose a meshless DR framework that operates on the parameter space of 3D Gaussian Splatting (3DGS). Our method employs two independent perturbation pipelines to synthesize randomized training datasets. First, a Photometric DR pipeline alters the baked illumination and color balance by modulating the Spherical Harmonics (SH) coefficients. Second, a Procedural DR pipeline isolates the subject's geometric shape by replacing its original textures with 3D spatial noise. Finally, these perturbed radiance fields are composited over stochastically varied backgrounds using a rasterization engine. Our parameter manipulation provides a meshless alternative for generating robust datasets for complex geometries.
\end{abstract}

\section{Introduction}
Generating synthetic data to close the Sim-to-Real gap traditionally relies on classical computer graphics rendering pipelines. While Domain Randomization (DR) is a standard technique for inducing robustness in these datasets, its implementation remains generally tied to polygonal meshes. However, 3D modeling and texturing of complex biological specimens, such as insects, are difficult.  These subjects consist of a mixture of 1D features (fine setae and antennae), planar 2D structures (semi-transparent wings), and volumetric 3D bodies. Forcing this mixed-dimensional geometry, along with complex transparencies and high-frequency textures, into a standard polygon mesh often requires manual effort and results in structural loss. This representational bottleneck is a primary reason meshless techniques such as 3D Gaussian Splatting (3DGS) have gained  traction~\cite{kerbl20233d}.

In biodiversity studies, automated identification of species suffers from a texture bias, where classification models rely heavily on local textural patterns rather than critical global morphological structures~\cite{geirhos2018imagenettrained, bjerge2023accurate}. Overcoming this bias presents a challenge for traditional mesh-based DR systems~\cite{tobin2017domain,blair2025leveraging}. In these frameworks, randomizing textures to encourage morphological generalization can degrade essential visual cues necessary for taxonomic identification \cite{margapuri2021classification}.

To bypass the limitations of traditional mesh-based 3D models, we propose a meshless DR framework based on 3DGS that represents scenes as unstructured collections of 3D anisotropic Gaussians. Operating within this continuous parameter space, our method applies two independent perturbation pipelines to synthesize randomized training datasets. First, a Photometric DR pipeline applies two distinct modulations to the Spherical Harmonics (SH) coefficients: it alters the overall color balance by shifting the base color components, and it modifies the baked directional illumination by rotating the primary lighting and scaling specular reflections. Second, a Procedural DR pipeline isolates the subject's morphology by replacing its original textures with 3D spatial noise. 
Bounding box extraction is essential for training detection architectures~\cite{deogan2025,USHIRO2026}. To address this, we eliminate the need for manual annotation by extracting 2D bounding boxes from the rendering engine.

By manipulating the radiance field through these pipelines, our method generates synthetic datasets that preserve the morphological integrity required to distinguish biological species, especially hyperdiverse and poorly understood groups~\cite{deep_darktaxa}. This approach can standardize the biological identification pipeline, enabling generalized training for species discovery and taxonomic classification.

The main contributions of our work are:
\begin{enumerate}
\item A two-stage, meshless DR pipeline combining offline 3DGS parameter manipulation with standalone game-engine rasterization to automate the generation of annotated datasets for species identification, bypassing the need for mesh-based surface extraction.
\item A Photometric DR pipeline that modulates SH via first-order Wigner D-matrices and stochastic scaling to synthesize illumination and color balance variations.
\item A Procedural DR pipeline that applies 3D spatial noise, preserving the underlying geometric primitives.
\end{enumerate}

\section{Related Work}
Domain Randomization exposes models to procedural variations in textures, lighting, and camera poses, bridging the Sim-to-Real gap~\cite{tobin2017domain}. Long established in robotics, these synthetic data and randomization frameworks are now advancing into biological domains~\cite{margapuri2021classification, aghamohammadi2025blender}. Translating these pipelines to complex real-world biological scenarios~\cite{bjerge2023accurate} reveals critical limitations. Existing biodiversity datasets, such as the 3D-scanned mammal collections~\cite{blair2025leveraging}, can rely on explicit Wavefront meshes paired with bitmap textures. These representations are effective for macroscopic diffuse structures but fail to accurately reproduce the complex light-surface interactions of insect morphology, such as high-frequency view-dependent specular reflections, spatially varying iridescence, and the directional transmittance of semi-transparent wings.

To overcome the limitations of polygonal modeling, 3D Gaussian Splatting (3DGS)~\cite{kerbl20233d} has emerged as an explicit, point-based radiance-field representation. While recent surveys highlight its rapid adoption across domains such as segmentation, generative modeling, and robotics~\cite{he2026survey,chen2026survey}, its application as a native, procedural data-synthesis engine remains largely unexplored. Recent efforts~\cite{USHIRO2026,deogan2025} have begun to use 3DGS models as static assets in game engines; however, this approach fails to fully exploit the potential of radiance-field parameterization. Conversely, to bridge 3DGS and established mesh-based DR pipelines, a common strategy is to extract explicit surfaces from these unstructured fields. For instance, methods such as SuGaR~\cite{guedon2024sugar} align flat Gaussians with the underlying geometry to facilitate mesh extraction. 
Yet, for organic specimens, surface extraction inherently degrades semi-transparent biological structures (e.g., setae) and permanently bakes view-dependent illumination into static textures.
This geometric conversion loss underscores the necessity of DR frameworks that operate directly on point-based radiance representations, avoiding the mesh-extraction bottleneck entirely.

The explicit parameterization of 3DGS has recently been identified as a controllable workspace for scene manipulation. Prior research has demonstrated the potential to influence Gaussian geometry for articulated movement~\cite{hu2024gauhuman,Zioulis2025Skinning} and to compute material and scene properties~\cite{liang2024gsir, Gao2025Relightable}. While these frameworks use parameter-space operations to recover feature properties or align geometry, our approach repurposes these perturbations to induce stochastic variability. By treating 3DGS primitives as a perturbable parameter space for photometric and procedural randomization, we generate diverse training distributions, bypassing the need for manual, engine-dependent material and lighting reconfiguration.

\section{Proposed Framework}
Our meshless DR pipeline automates the synthesis of annotated biological datasets by operating directly on the continuous parameter space of 3D Gaussian Splatting. The workflow consists of three sequential stages. First, a reconstructed 3DGS model undergoes spatio-optical pruning to eliminate geometric artifacts. Second, the filtered radiance field is subjected to two independent stochastic perturbation branches: a Photometric pipeline that modulates illumination and color balance, and a Procedural pipeline that applies 3D spatial noise to isolate morphology. Finally, these mathematically perturbed models are ingested by a custom Unity rasterization engine. This engine acts as an automated virtual studio, compositing the specimens over randomized backgrounds and analytically extracting zero-cost 2D annotations across arbitrary camera poses.

\subsection{Explicit Parameterization and Pruning}
Let us consider a 3DGS model with $N$ individual Gaussian primitives with geometric and volumetric attributes, position $\mu \in \mathbb{R}^{N \times 3}$, anisotropic scale $\mathbf{s} \in \mathbb{R}^{N \times 3}$, rotation quaternions $\mathbf{q} \in \mathbb{R}^{N \times 4}$, and opacity $\alpha \in \mathbb{R}^{N}$. Following Kerbl~et~al.~\cite{kerbl20233d}, the view-dependent optical properties are modeled by Spherical Harmonics (SH) coefficients $\mathbf{c} \in \mathbb{R}^{N \times 16 \times 3}$. 

To execute perturbations directly within this multidimensional space, we categorize the SH parameters based on their degree $\ell$. The degree $\ell=0$ represents the base view-independent color (analogous to intrinsic albedo), while higher degrees ($\ell>0$) capture high-frequency, view-dependent directional reflections. 

Before applying the stochastic pipelines, we execute a \emph{spatio-optical pruning step} to refine the baseline model, inspired by structural filtering techniques used to mitigate floaters and rendering artifacts~\cite{kerbl20233d, wang20243dgaa, hu2024gauhuman}. Although low-opacity and highly elongated splats are often visually imperceptible in the native reconstruction, they pose a risk during domain randomization. When stochastic operations, such as aggressive color shifting or procedural noise substitution, are applied, these latent artifacts can be amplified, creating visual noise that degrades the subject's silhouette and structural details. To prevent this deterioration, splats are filtered out if they exhibit both opacity below 10\% and extreme spatial elongation, defined as the longest principal axis exceeding 10 times the shortest. These specific threshold values serve as empirical hyperparameters that can be tuned on a per-dataset basis. Furthermore, this stage is optional and may be bypassed entirely if the underlying 3DGS models are artifact-free.

\subsection{Pipeline A: Photometric Domain Randomization}
We compute two SH operators that decouple the diffuse color balance from view-dependent specular reflections.

\textbf{1) Photometric Color Shift:} Inspired by photometric modification and corruption models used in adversarial robustness \cite{wang20243dgaa}, we simulate sensor color balances and lighting variations by applying a global scaling factor $\lambda$ and a color shift $\mathbf{\Delta}_{color}$ to the DC component ($\ell=0$):
\begin{equation}
    \hat{\mathbf{c}}_0^0 = \lambda \mathbf{c}_0^0 + \mathbf{\Delta}_{color},
\end{equation}
where $\mathbf{c}_0^0 \in \mathbb{R}^3$ represents the original DC coefficients of the Gaussian primitive, and $\lambda \in \mathbb{R}$ is a stochastic scalar factor simulating global exposure variations. In our framework, the perturbation vector is sampled from a uniform distribution over the unconstrained SH space, such that $\mathbf{\Delta}_{color} \in [-1.2, 1.2]^3$. This specific bounded range serves as an empirical hyperparameter tuned for biological specimens; because these offsets are applied directly to the raw parameters prior to the rasterizer's projection and clamping stages, they enable a high-variance simulation of extreme white-balance shifts while preserving local structural gradients from premature pixel saturation.

\textbf{2) AC Wigner Rotation and Specular Scaling:} To simulate the rotation of primary light sources, we adapt SH rotation concepts explored in articulated avatars \cite{hu2024gauhuman, Zioulis2025Skinning} for static geometries. While rotating all SH bands via Wigner D-matrices is mathematically rigorous \cite{Zioulis2025Skinning}, applying these complex transformations to higher-order coefficients yields diminishing visual benefits given their high computational cost \cite{hu2024gauhuman}. Therefore, to maintain an optimal trade-off between visual quality and rendering throughput for data synthesis, we restrict the explicit lighting rotation to the primary directional lighting (degree $\ell=1$), where the Wigner D-matrix simplifies to a standard $3 \times 3$ Cartesian rotation $\mathbf{R}$~\cite{Zioulis2025Skinning}. Higher-order AC components ($\ell > 1$) bypass this rotation and are solely modulated by a stochastic log-normal specular scaling factor $\gamma_{spec}$. The updated AC coefficients are computed as:

\begin{equation}
\hat{\mathbf{c}}_\ell = 
\begin{cases} 
\gamma_{spec} \mathbf{R} \, \mathbf{c}_\ell & \text{for } \ell = 1 \\
\gamma_{spec} \mathbf{c}_\ell & \text{for } \ell > 1
\end{cases}.
\end{equation}

\subsection{Pipeline B: Procedural DR}
To achieve structural randomization and ensure that no view-dependent textural cues propagate into the procedural representation, the DC color component is replaced while all higher-order AC components are set to zero:

\begin{equation}
\hat{\mathbf{c}}_\ell = 
\begin{cases} 
\Phi(N(\mu_i)) \cdot \mathbf{C}_{amp} + \mathbf{C}_{shift} & \text{for } \ell = 0 \\
0 & \text{for } \ell > 0
\end{cases},
\end{equation}
where $N(\mu_i)$ represents a multi-scale 3D spatial noise function evaluated at the mean position $\mu_i$ of the $i$-th Gaussian. To encourage a diverse range of textural patterns, the transformation function $\Phi(\cdot)$ stochastically applies either a step function with a randomized threshold, binarizing the noise into discrete structural patches, or an identity mapping that preserves continuous color gradients. Finally, the dynamically scaled parameters $\mathbf{C}_{amp}$ and $\mathbf{C}_{shift}$ linearly map the resulting patterns into randomized color spaces.

\subsection{Game-Engine Rasterization and Compositing}
The generated sets of randomized 3DGS parameter tensors are imported into a custom Unity application running a dedicated 3DGS rasterization pipeline~\cite{pranckevicius2023unity}. 

This environment serves as an automated virtual studio, by sequentially positioning a virtual sensor at arbitrary poses around the asset. The engine then composites the rendered 3D biological specimens over a diverse array of stochastically sampled 2D background textures mapped onto planes, terrains, and skyboxes. 

During the rasterization phase, the 2D bounding box is extracted by projecting the model's 3D bounding box onto the image plane.  

\section{Results}
\begin{figure}[htpb]
    \centering
    \setlength{\tabcolsep}{0pt}
    \renewcommand{\arraystretch}{0}

    \begin{tabular}{cc}
        \includegraphics[width=0.5\linewidth]{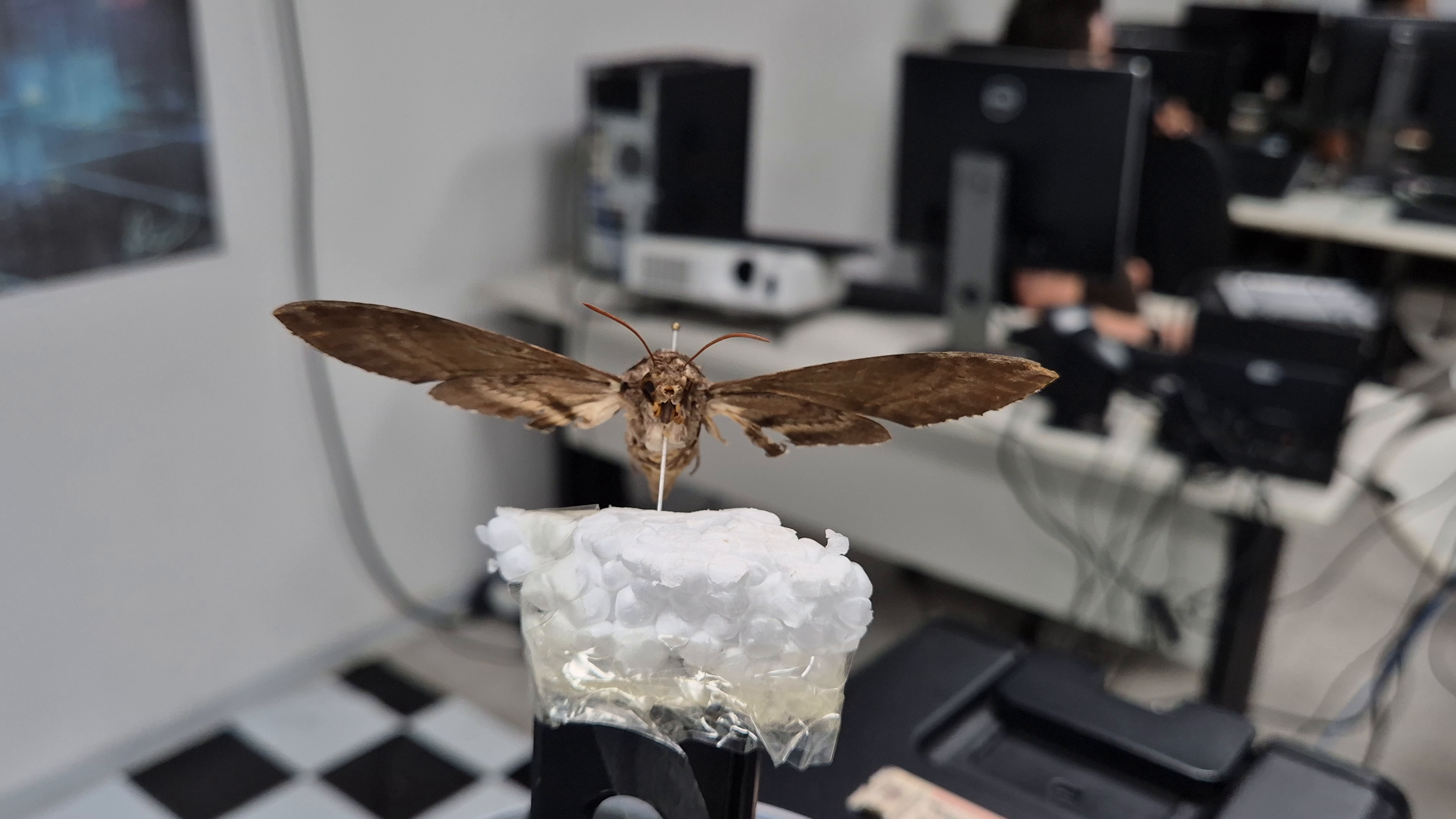}&
        \includegraphics[width=0.5\linewidth]{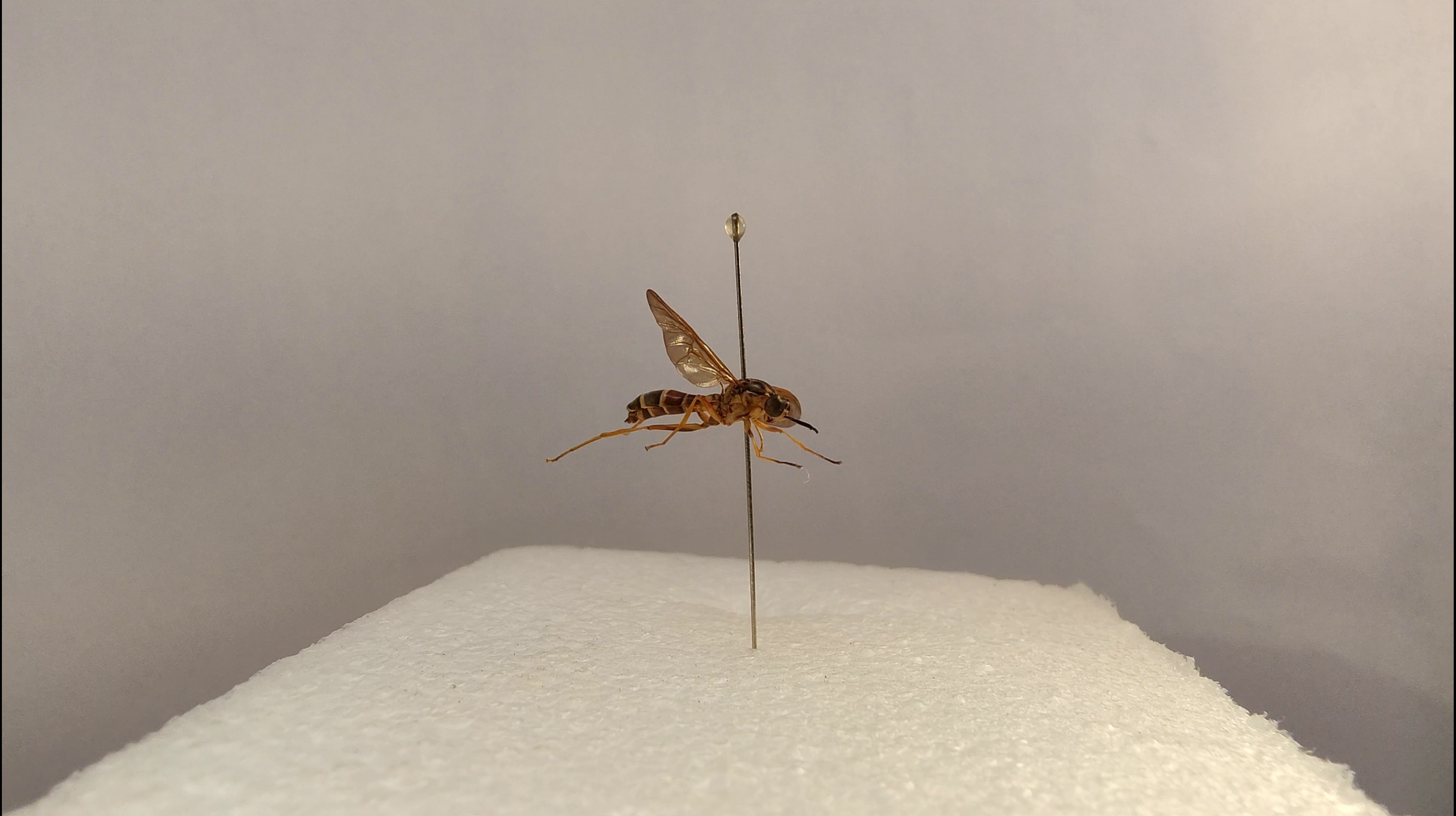}\\
        \includegraphics[width=0.5\linewidth]{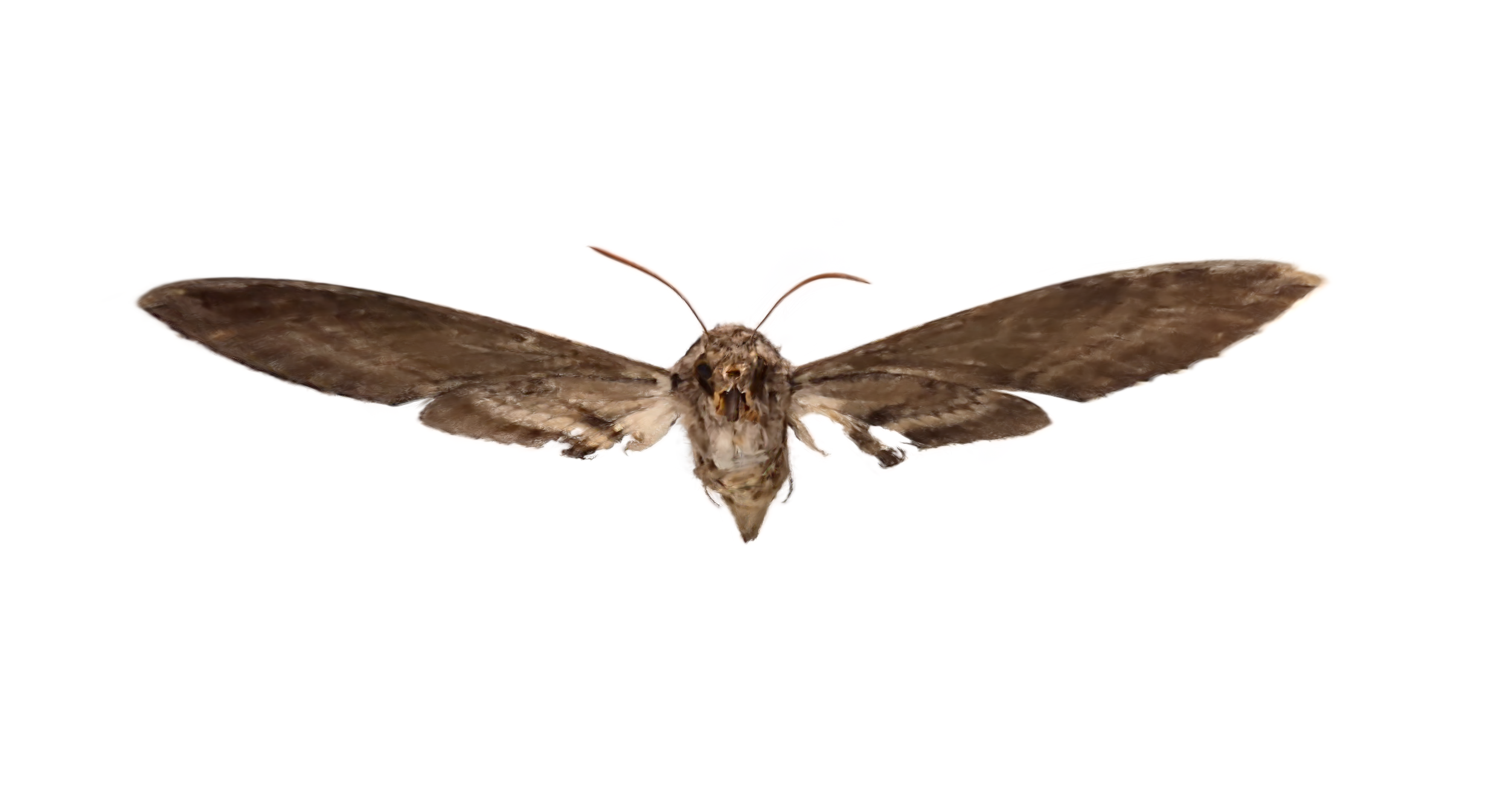}&
        \includegraphics[width=0.5\linewidth]{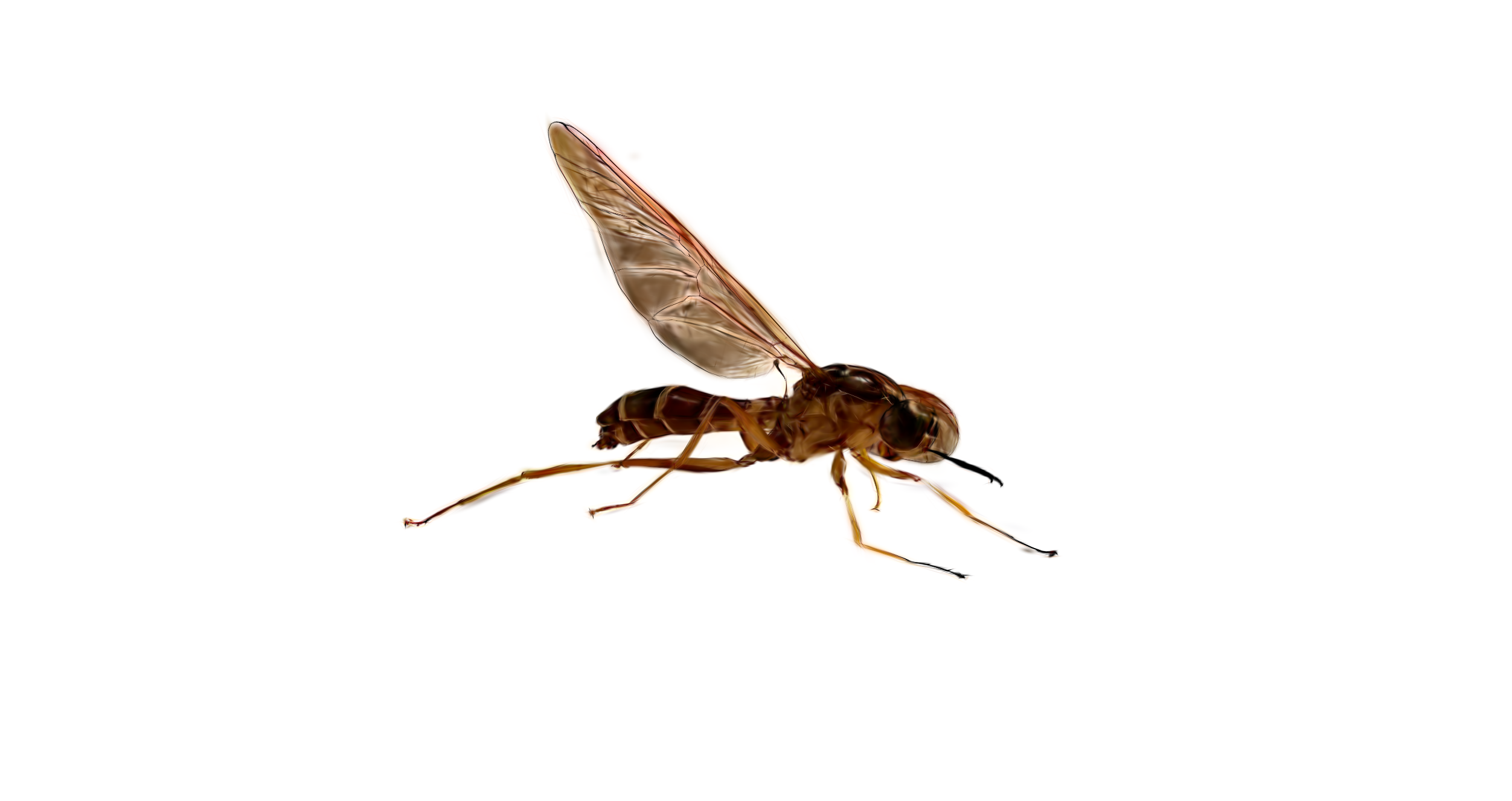}
    \end{tabular}

\caption{Baseline 3DGS reconstructions. Top: Representative free-hand multi-view video frames of a moth and a fly captured via mobile devices. Bottom: The corresponding canonical 3D Gaussian Splatting models, manually cropped to isolate the morphological structures prior to parameter perturbation.}
\label{fig:real3dgs}
\end{figure}

To obtain the baseline 3D scene representations, casual multi-view video sequences of the biological specimens were captured using standard mobile devices (a Motorola Edge 30 Fusion for the fly specimen and a Samsung Galaxy S24 for the moth). The initial 3D models were reconstructed using the off-the-shelf implementations of SuGAR~\cite{guedon2024sugar} (bypassing the optional mesh extraction stage to retain the pure splat representation) and EFA-GS~\cite{wang2025lowfrequency}. Because our core contribution lies in the domain randomization of radiance fields rather than in the reconstruction phase itself, we relied on the default optimization parameters provided by these established frameworks. The raw 3DGS models were subsequently imported into the SuperSplat~\cite{supersplat} editor, where manual spatial cropping was performed to isolate the specimens' morphology by removing background geometry and peripheral floaters (Fig.~\ref{fig:real3dgs}).

After manual isolation, we applied our spatio-optical pruning step to sanitize the baseline models. Initially, the moth model consisted of 154330 splats, and the fly model had 15413 splats. After removing low-opacity and highly elongated artifacts, the models were reduced to 95406 and 9166 splats, respectively, yielding cleaner geometric priors.

\begin{figure*}[htpb]
    \centering
    \setlength{\tabcolsep}{0pt}
    \renewcommand{\arraystretch}{0}
    
    \begin{tabular}{cccc}
        \includegraphics[width=0.245\linewidth]{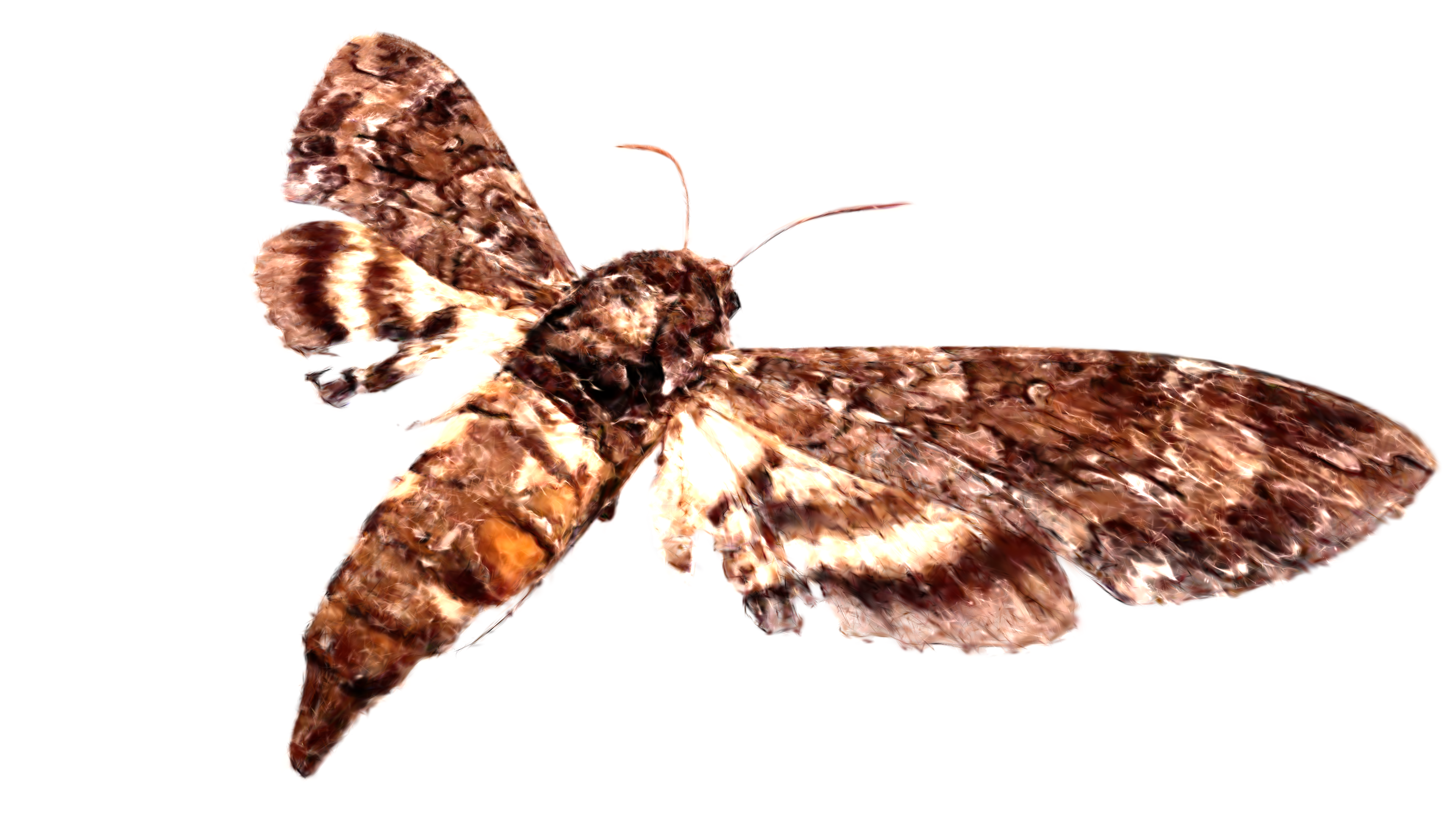} &
        \includegraphics[width=0.245\linewidth]{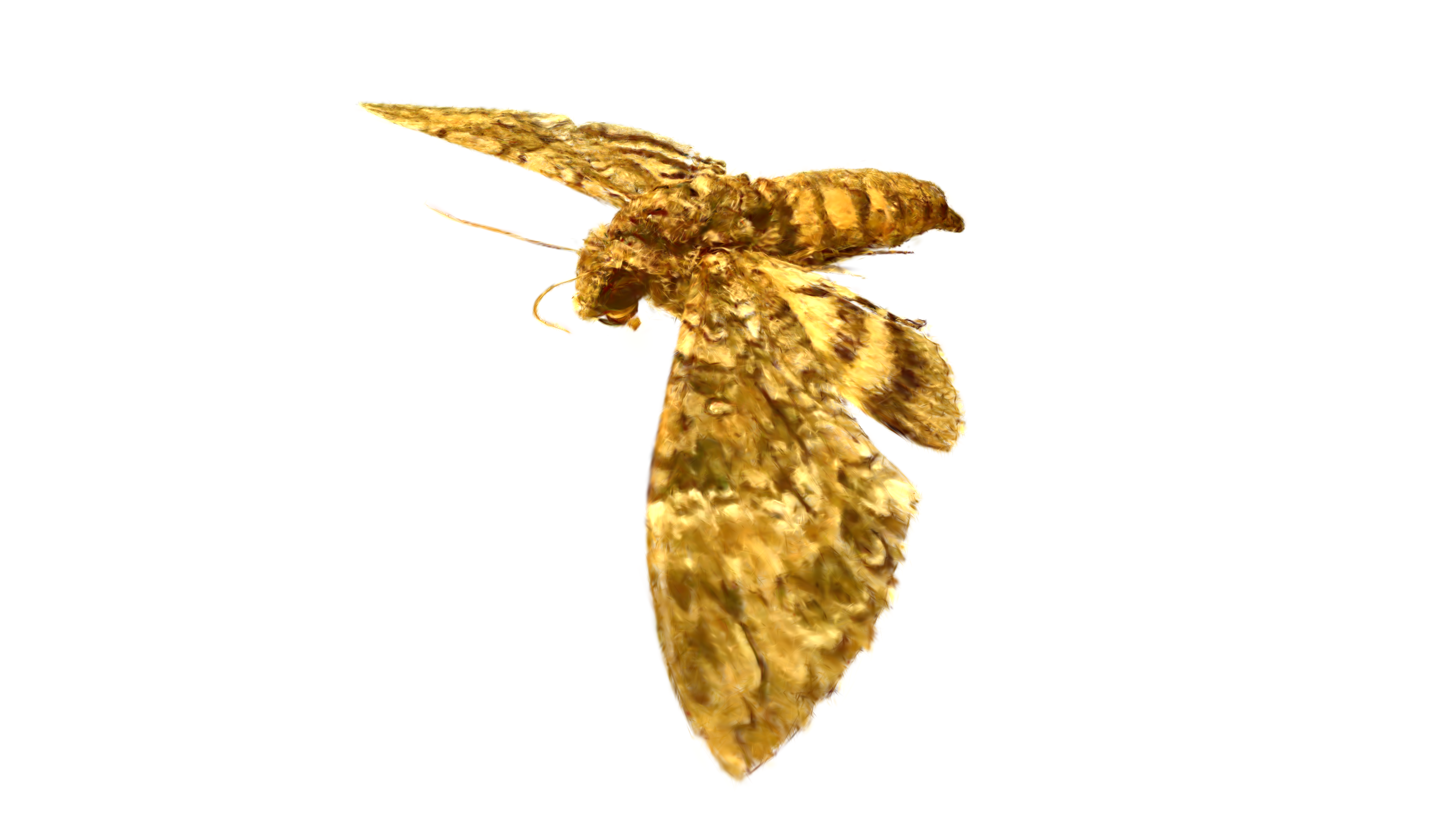} &
        \includegraphics[width=0.245\linewidth]{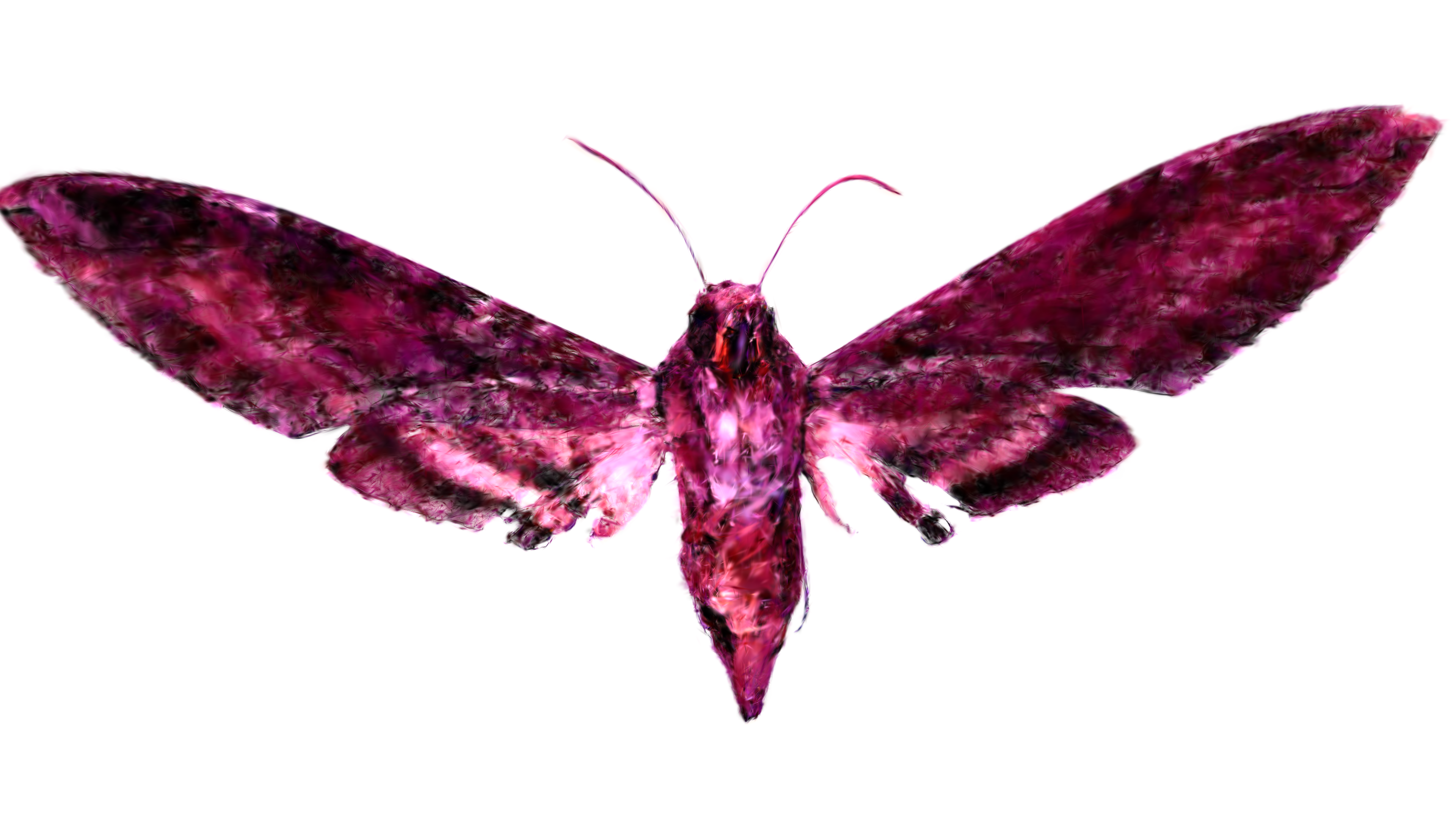} &
        \includegraphics[width=0.245\linewidth]{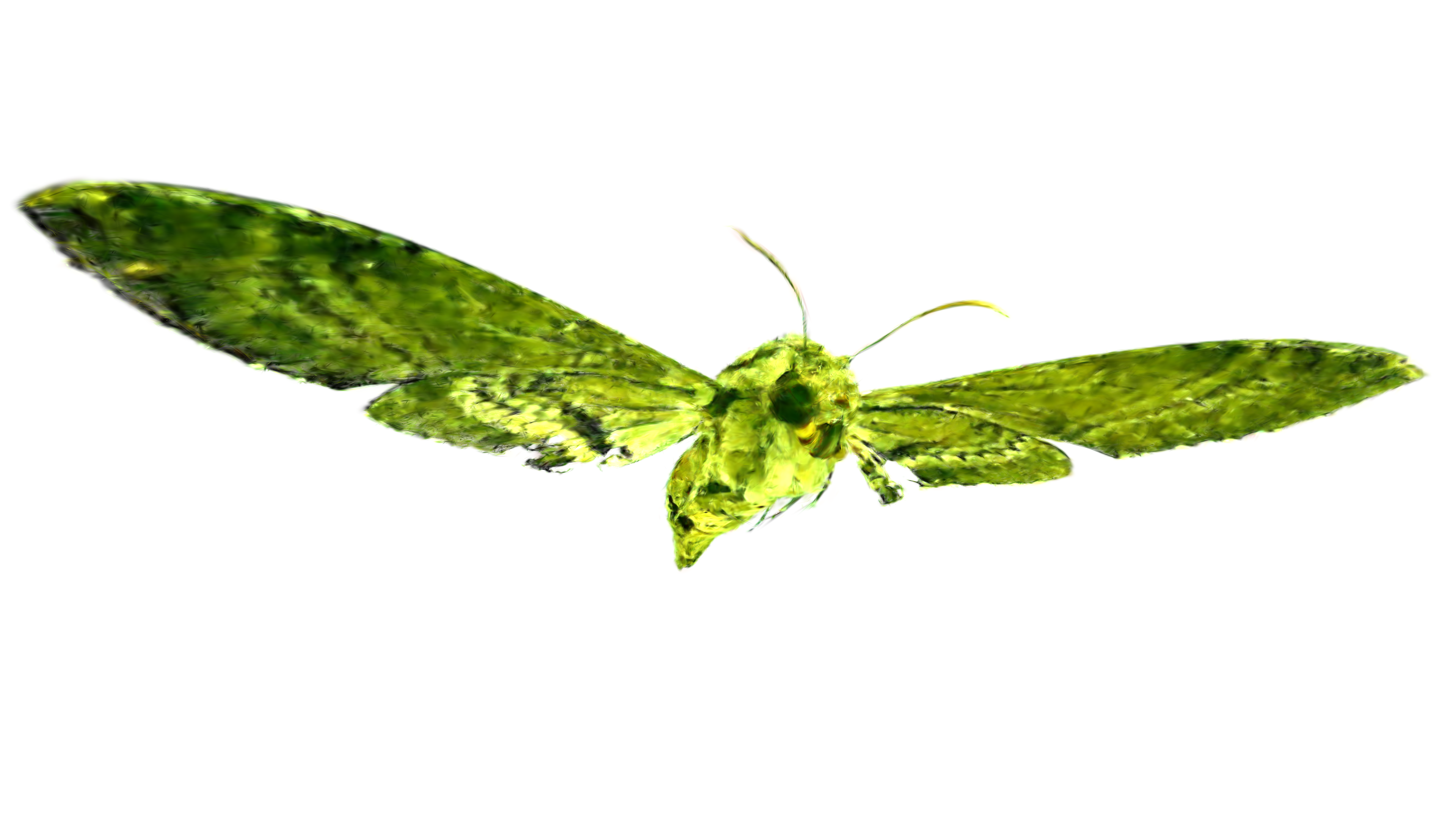} \\
        
        \includegraphics[width=0.245\linewidth]{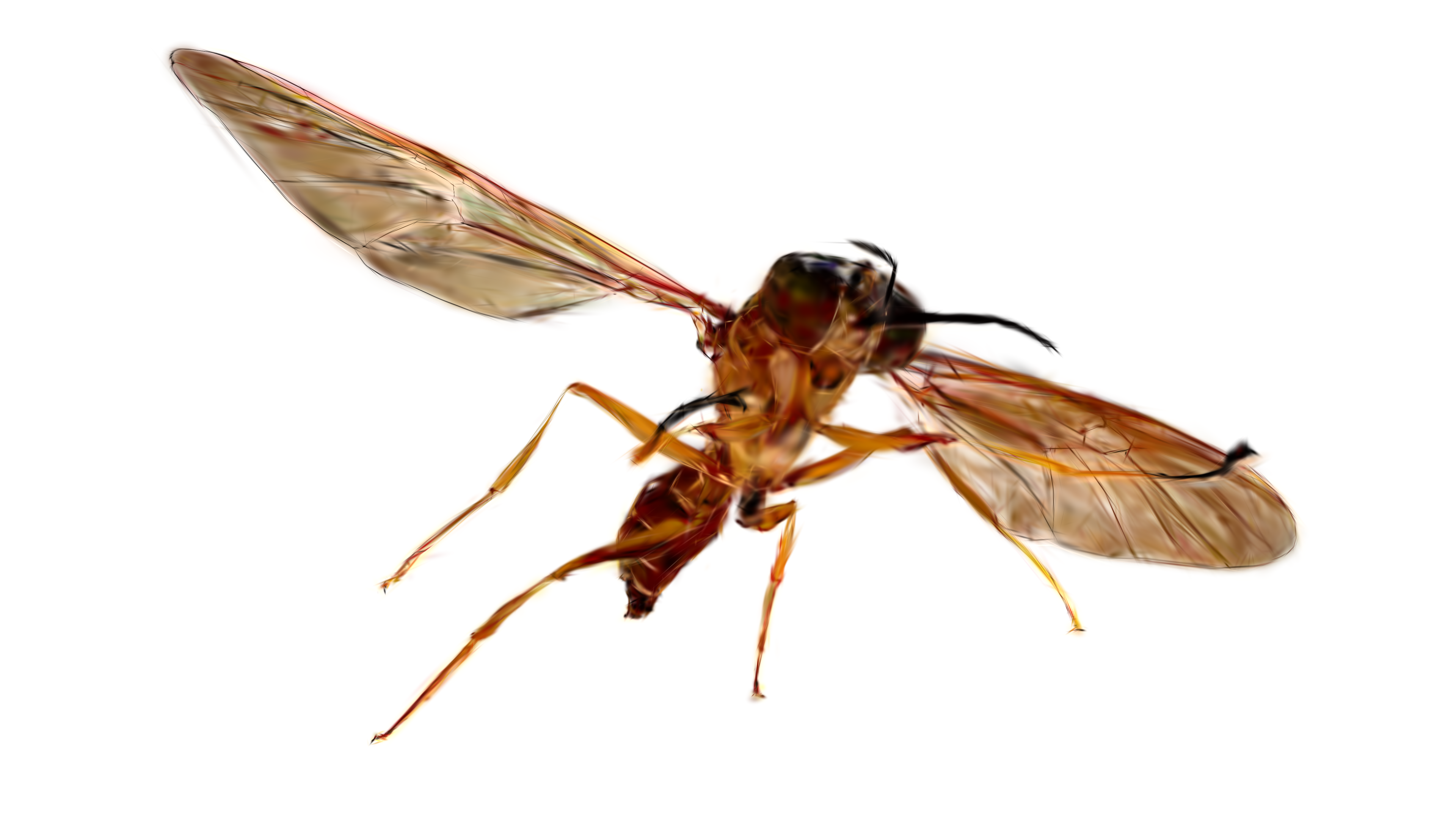} &
        \includegraphics[width=0.245\linewidth]{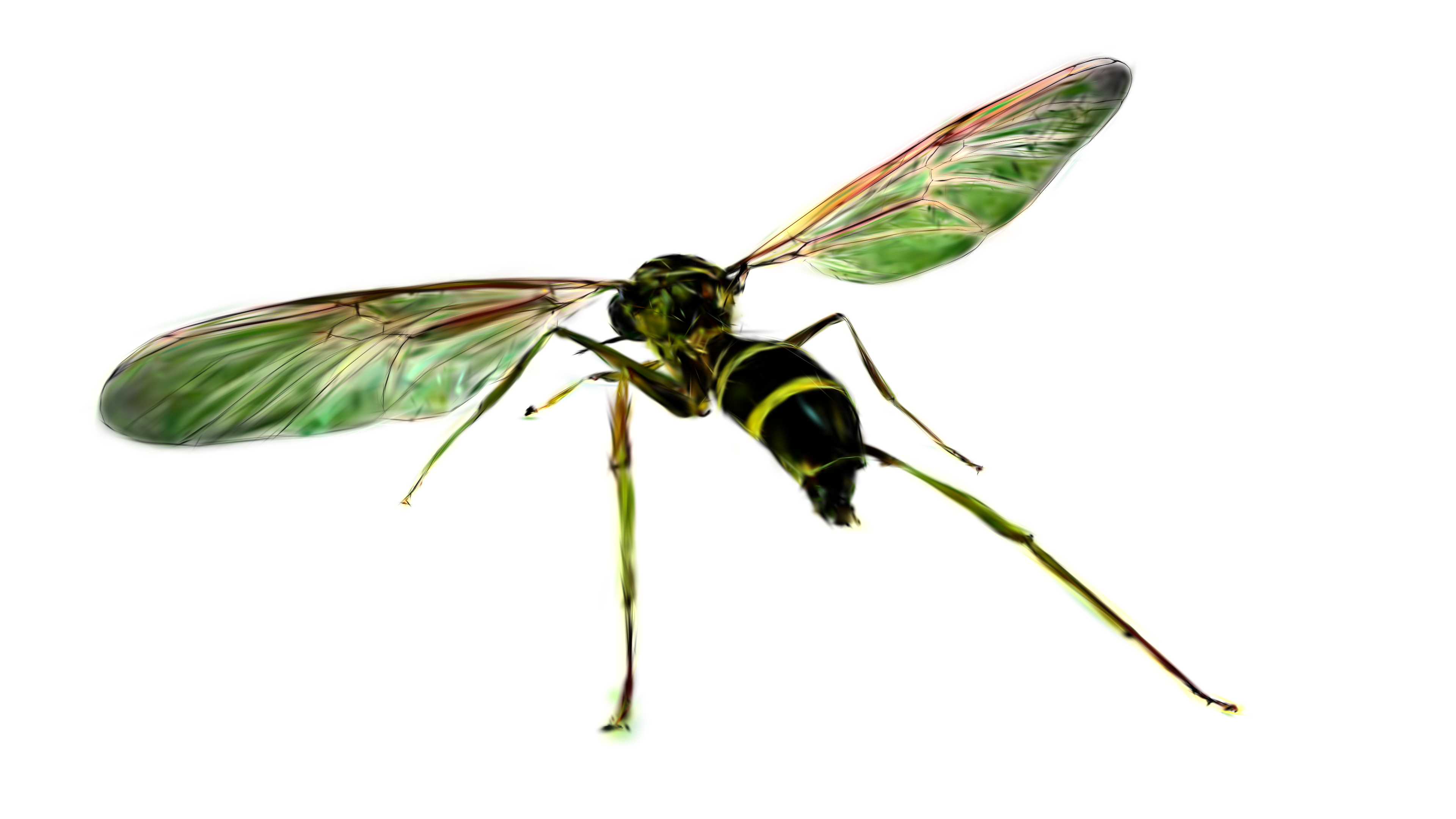} &
        \includegraphics[width=0.245\linewidth]{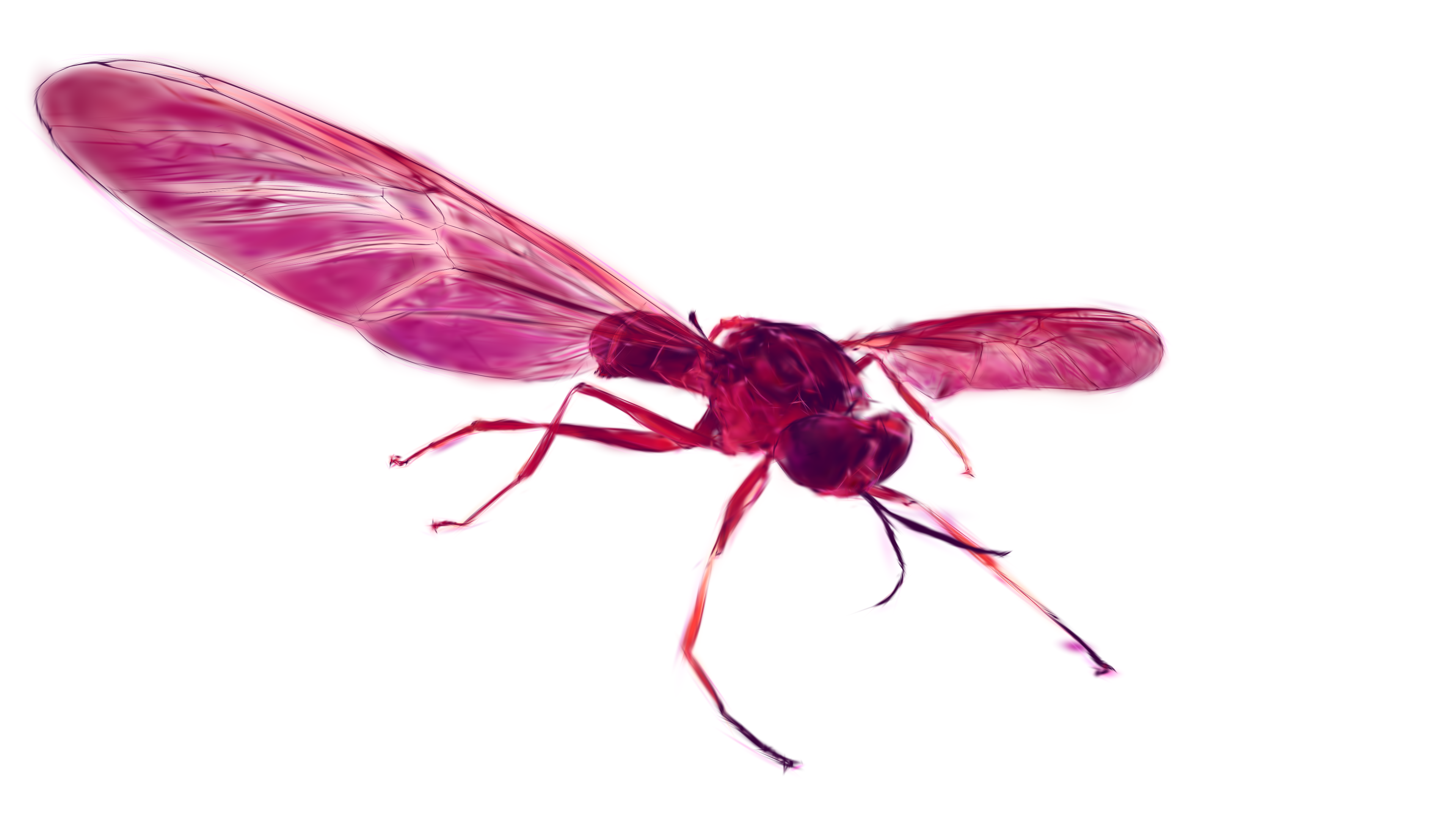} &
        \includegraphics[width=0.245\linewidth]{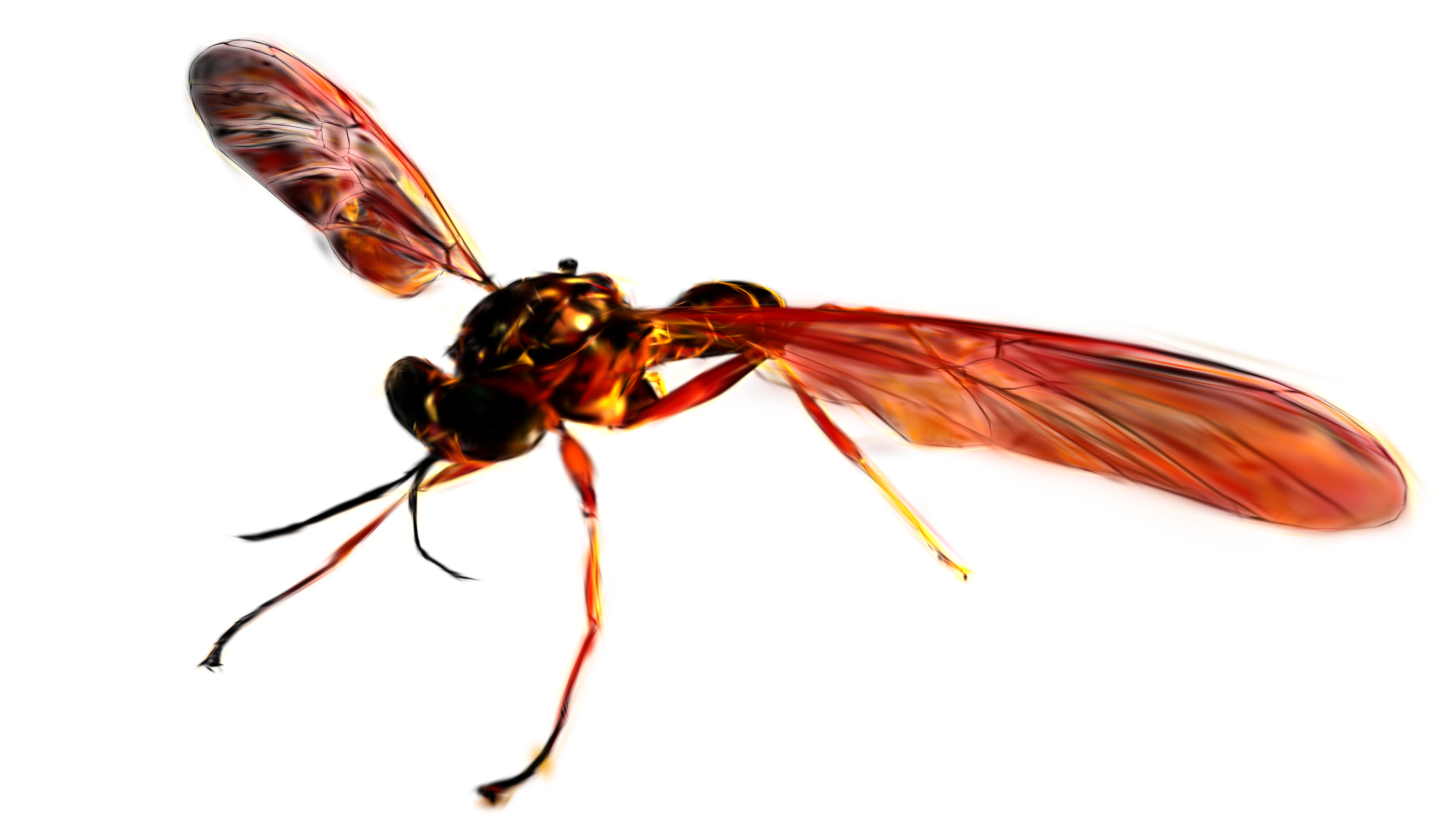} \\
        
    \end{tabular}
        
    \caption{Photometric DR pipeline outputs applied to the moth and fly 3DGS models.}

    \label{fig:photometric}
\end{figure*}

\begin{figure*}[htpb]
    \centering
    \setlength{\tabcolsep}{0pt}
    \renewcommand{\arraystretch}{0}
    
    \begin{tabular}{cccc}
 
\includegraphics[width=0.225\linewidth]{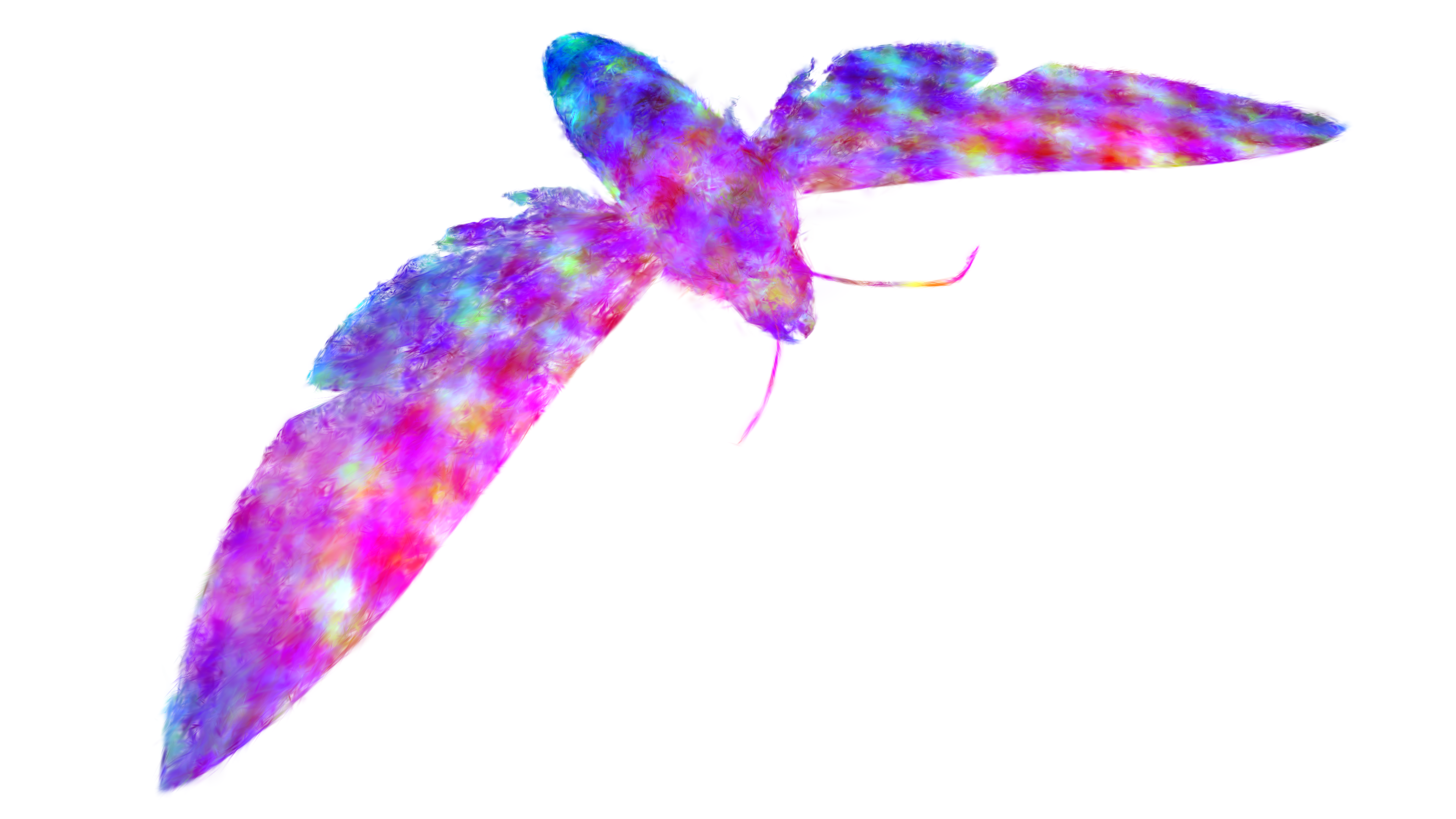} &
\includegraphics[width=0.225\linewidth]{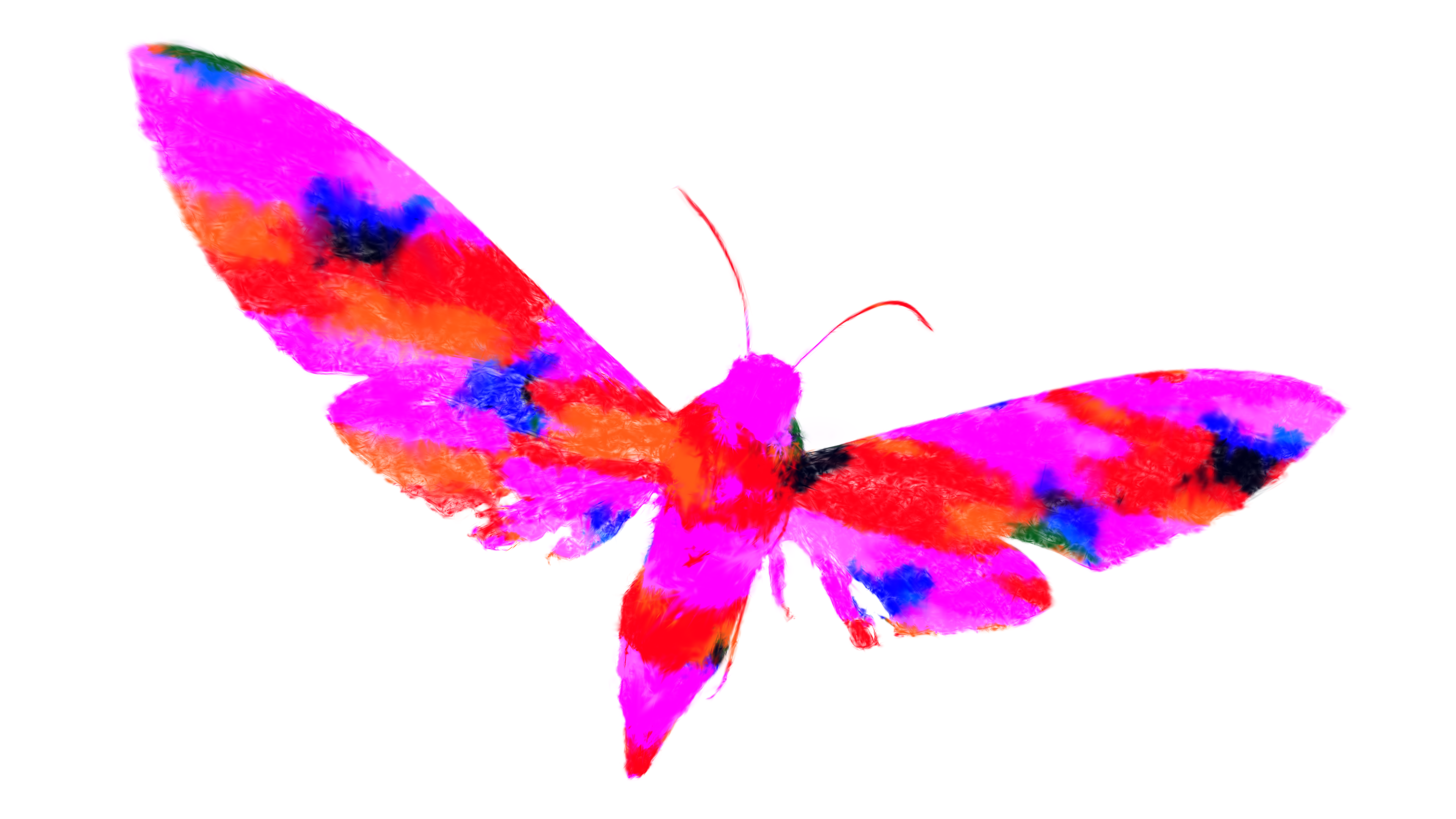} &
\includegraphics[width=0.225\linewidth]{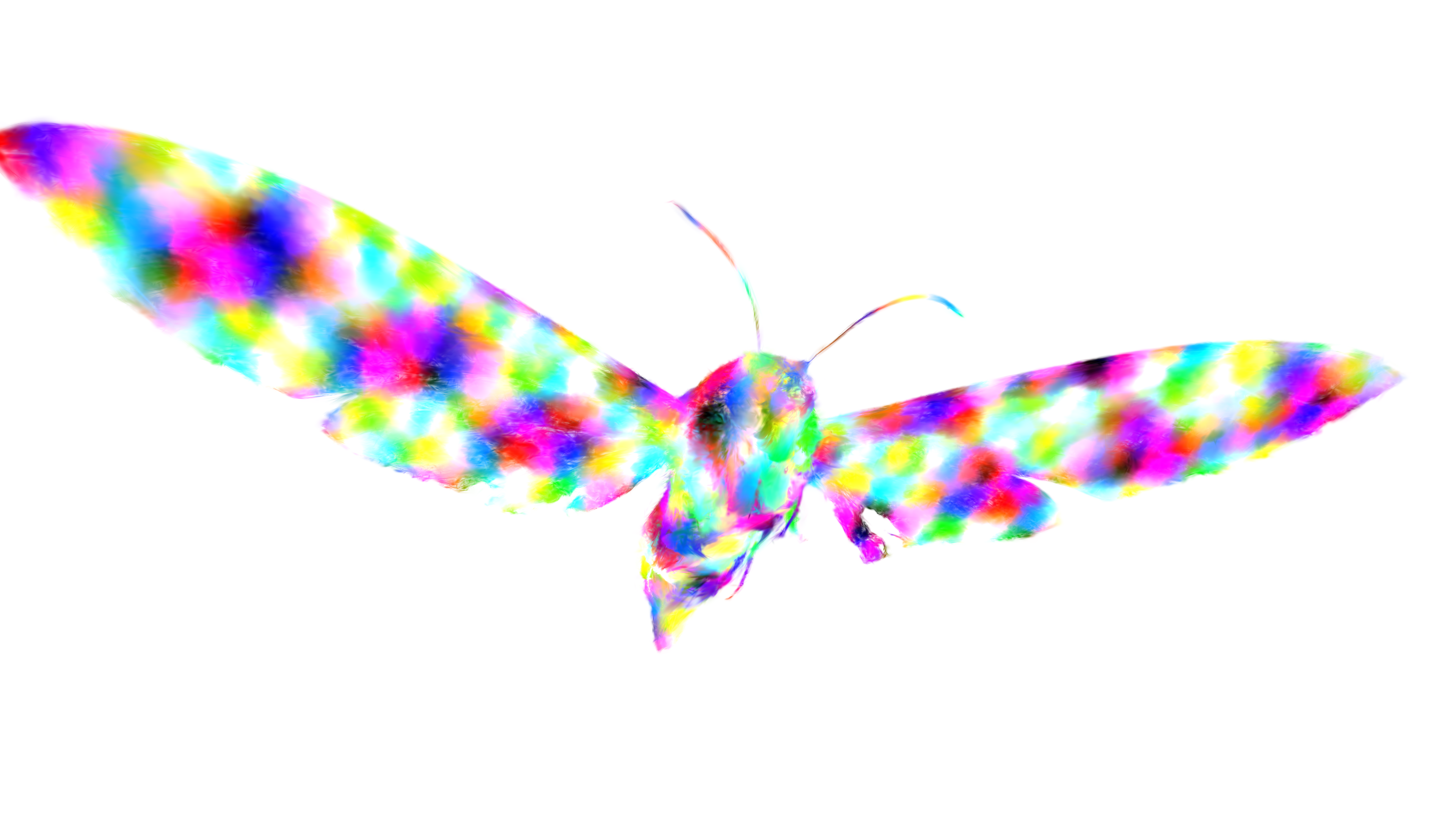} &
\includegraphics[width=0.225\linewidth]{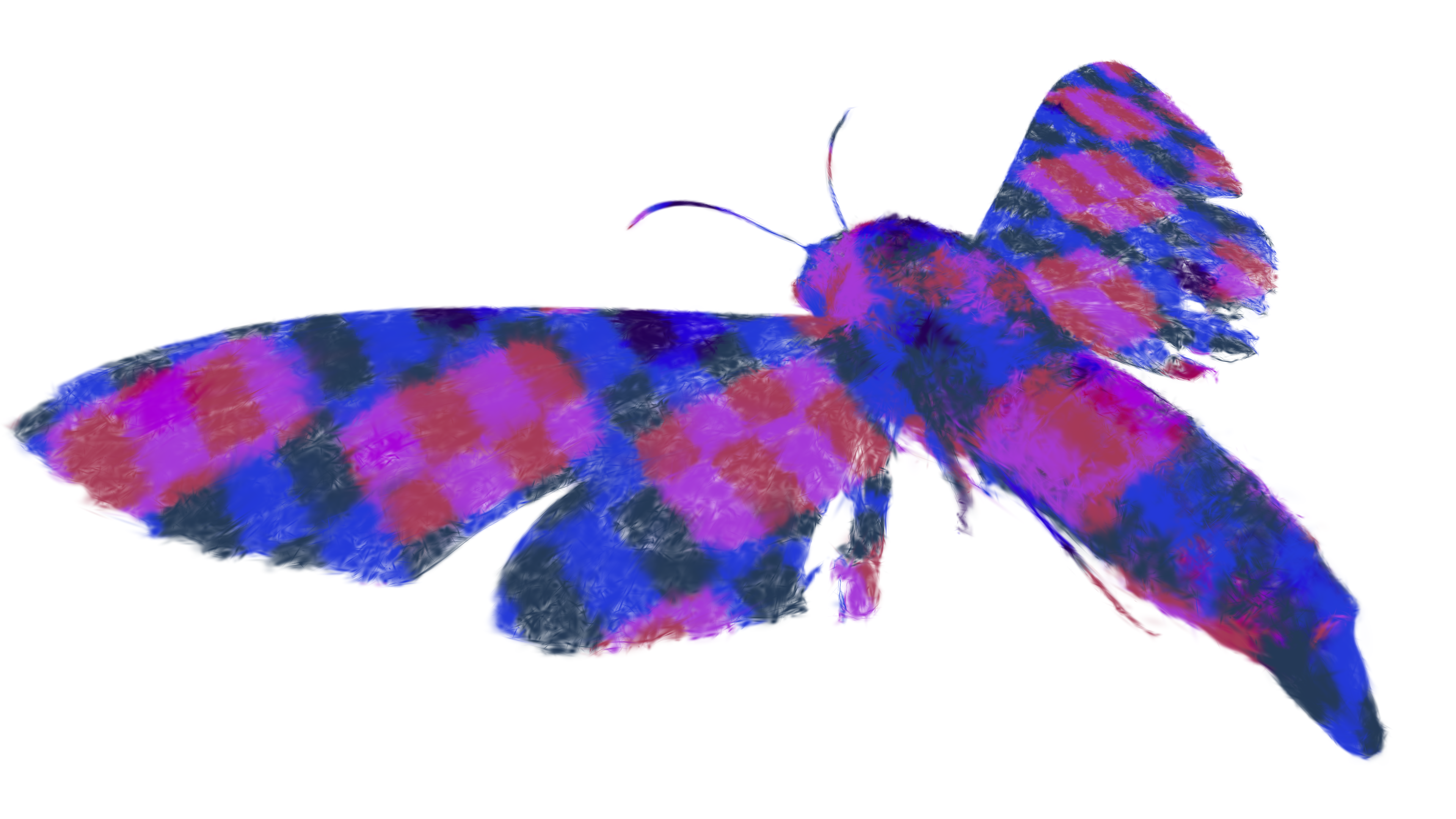} \\

\includegraphics[width=0.225\linewidth]{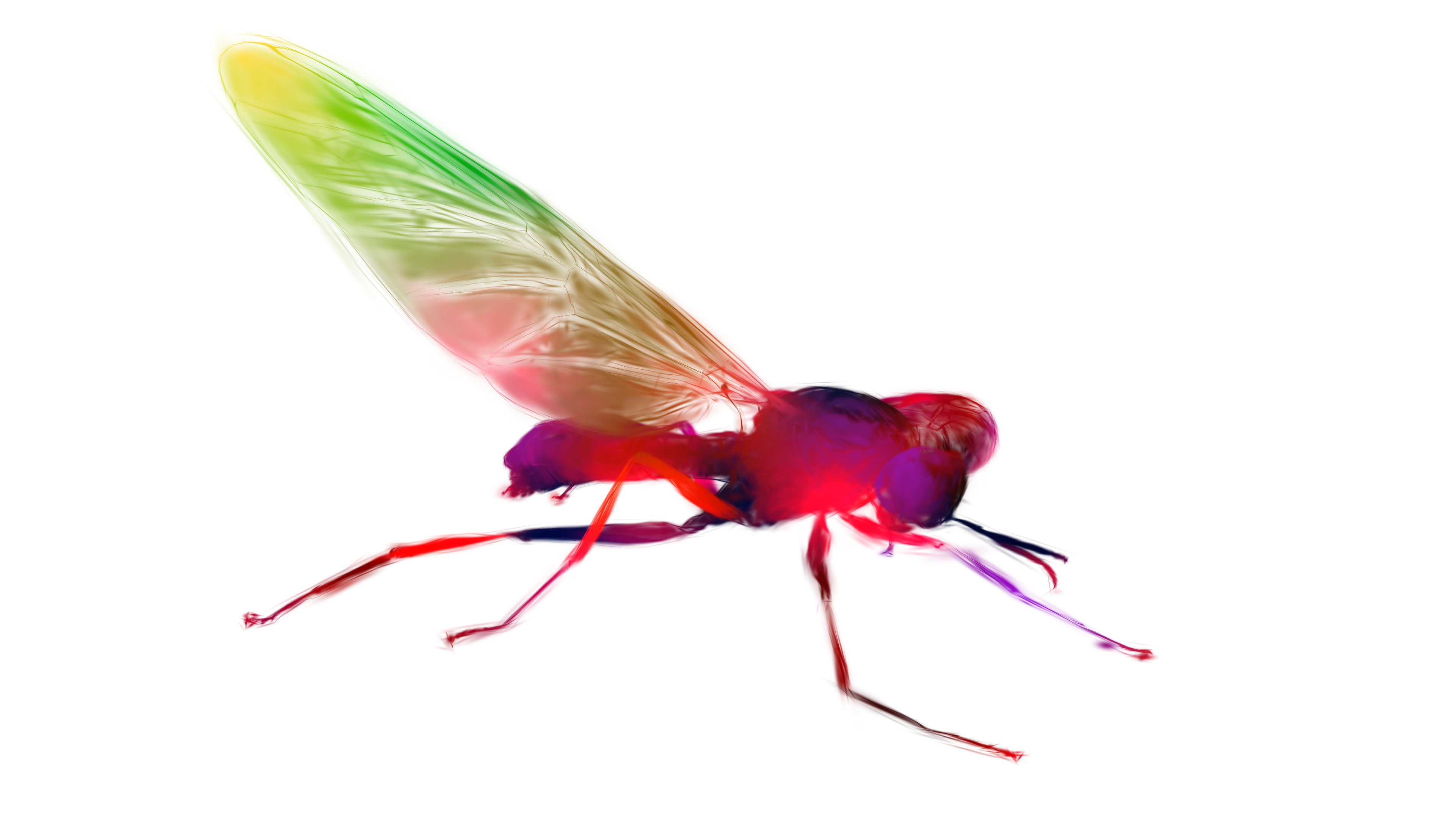} &
\includegraphics[width=0.225\linewidth]{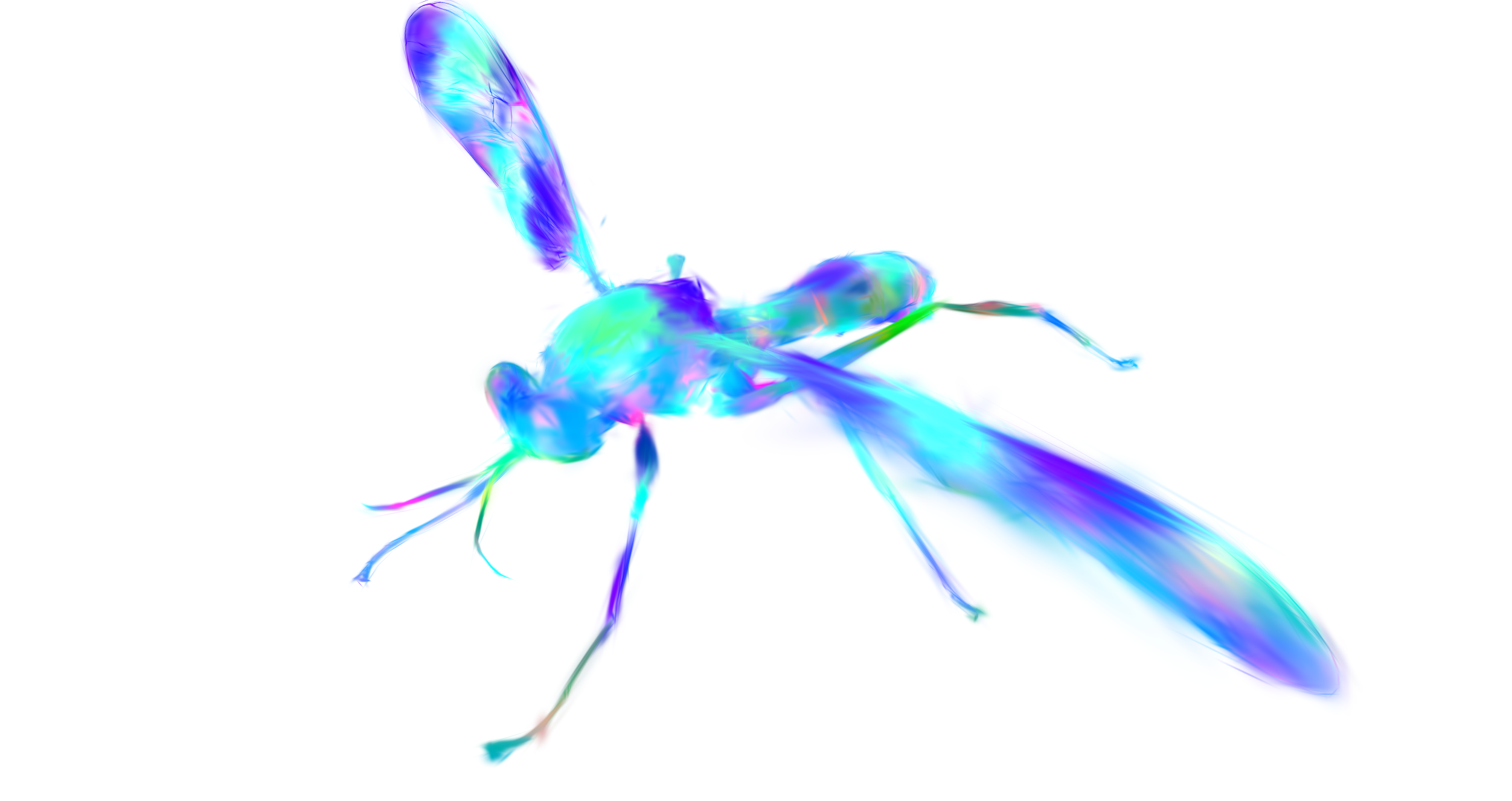} &
\includegraphics[width=0.225\linewidth]{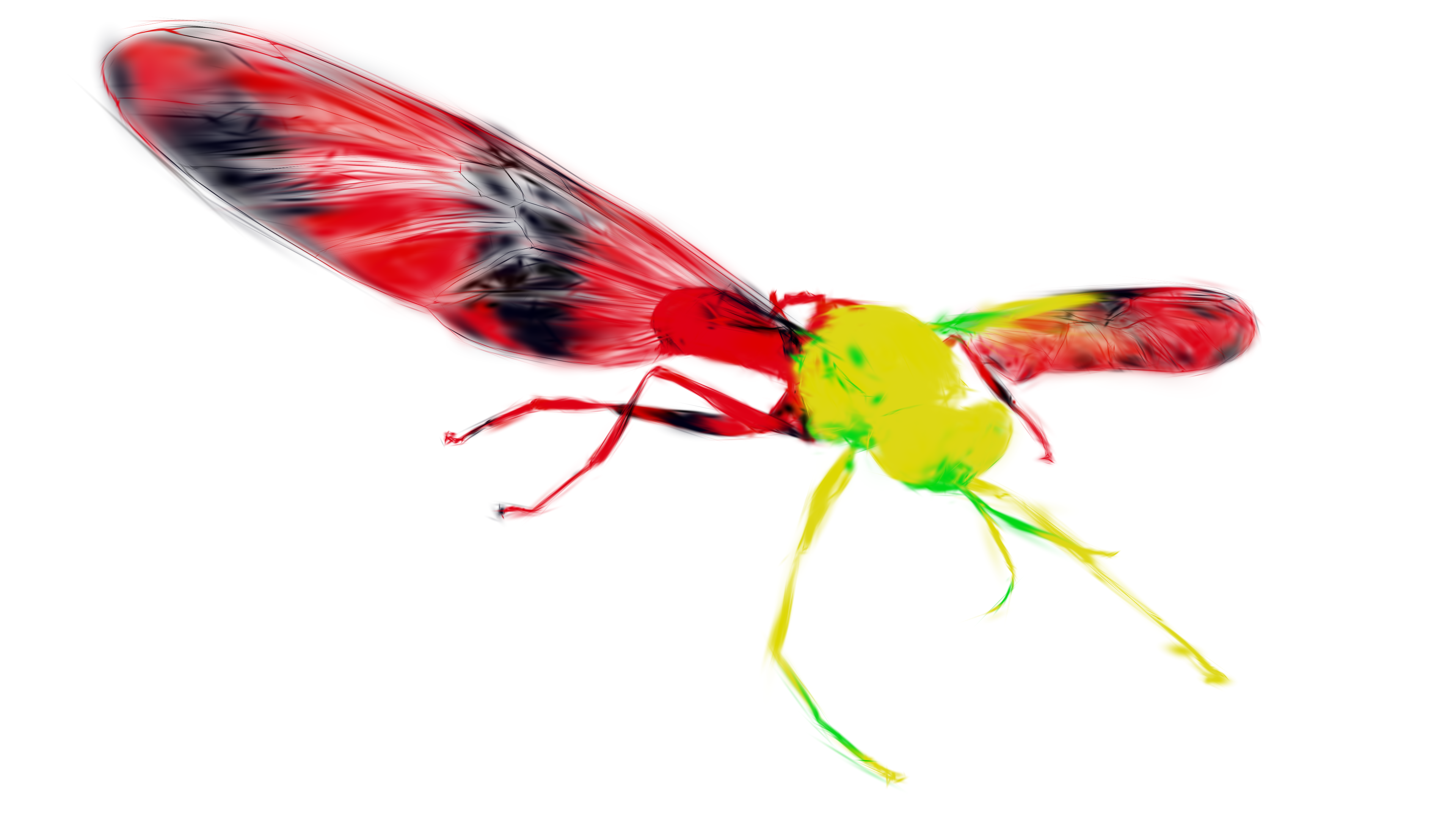} &
\includegraphics[width=0.225\linewidth]{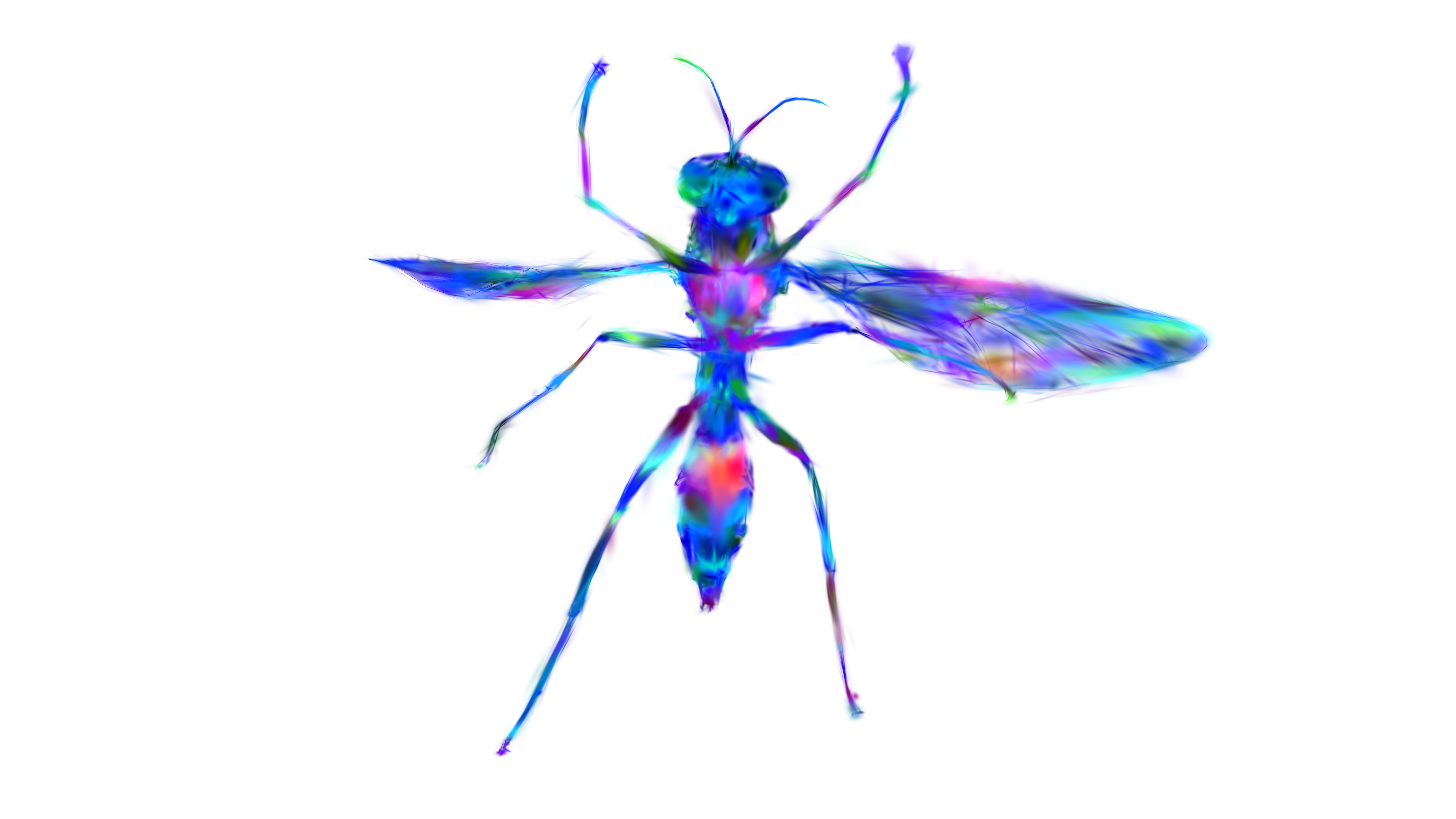}
    \end{tabular}
    
    \caption{Procedural DR pipeline outputs. Native textures are overridden by 3D spatial noise mappings, decoupling shape from appearance.}
        
    \label{fig:procedural}
\end{figure*}

\subsection{Feature Space Domain Coverage}
To show that our perturbation pipelines expand the visual manifold beyond the original captures, we conducted a feature-space analysis. We extracted high-dimensional deep feature embeddings using a pre-trained ResNet-50 architecture~\cite{HeResNet} and projected them into 2D using UMAP~\cite{mcinnes2018umap}. The evaluation dataset was structured into four distinct categories of 100 images each: (1) segmented original real-world video frames with backgrounds removed; (2) unperturbed control renderings from the canonical 3DGS model; (3) randomized renderings generated by our DR pipelines against a neutral background; and (4) randomized renderings composited over stochastic backgrounds. To ensure a fair baseline fidelity, the canonical control images were rendered from manually aligned camera poses that closely matched original trajectories.

\begin{figure}[ht!]
    \centering
    \begin{subfigure}[b]{0.48\textwidth}
        \centering
        \includegraphics[width=\textwidth]{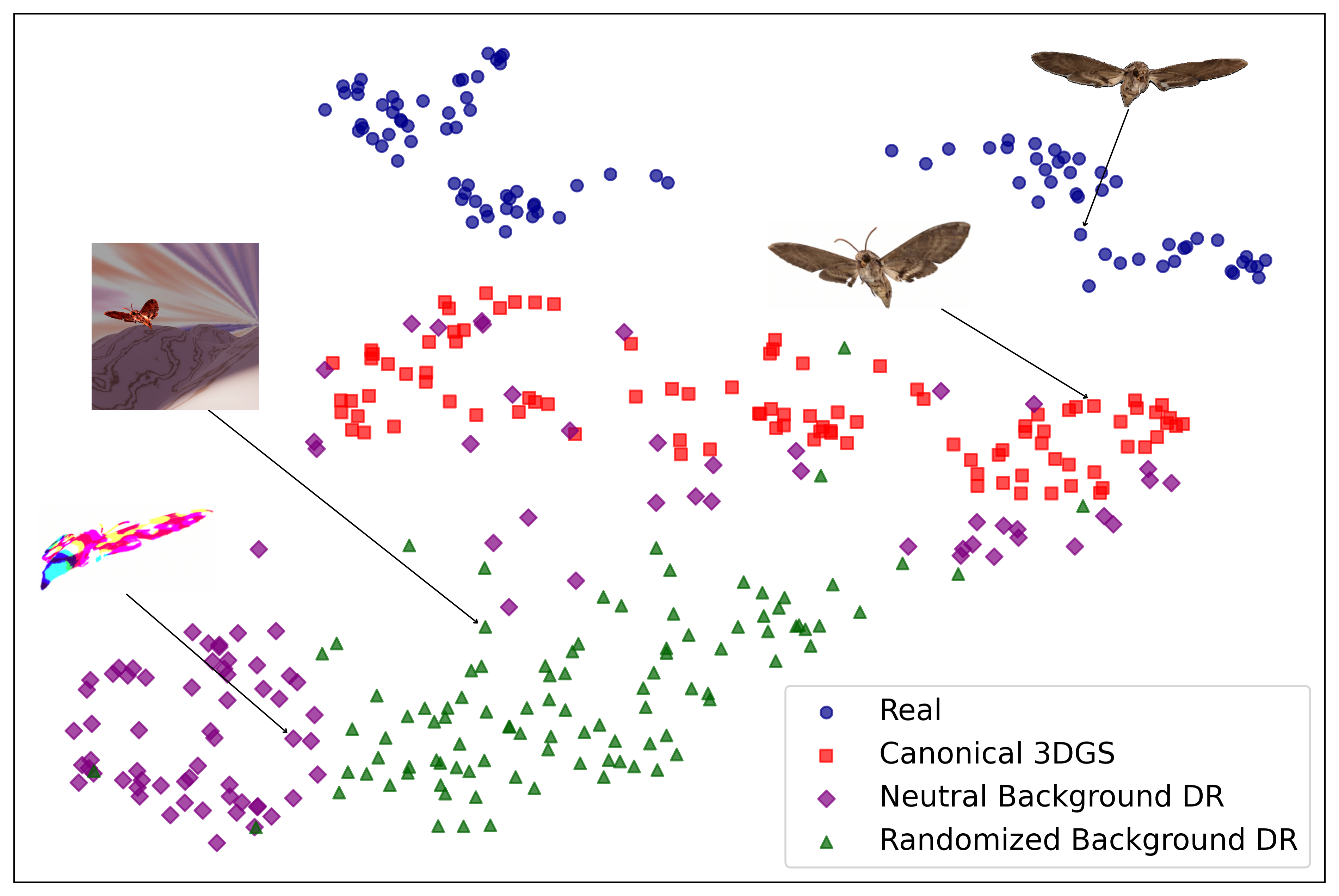}
        \caption{Moth specimen}
        \label{fig:umap_moth}
    \end{subfigure}
    \hfill
    \begin{subfigure}[b]{0.48\textwidth}
        \centering
        \includegraphics[width=\textwidth]{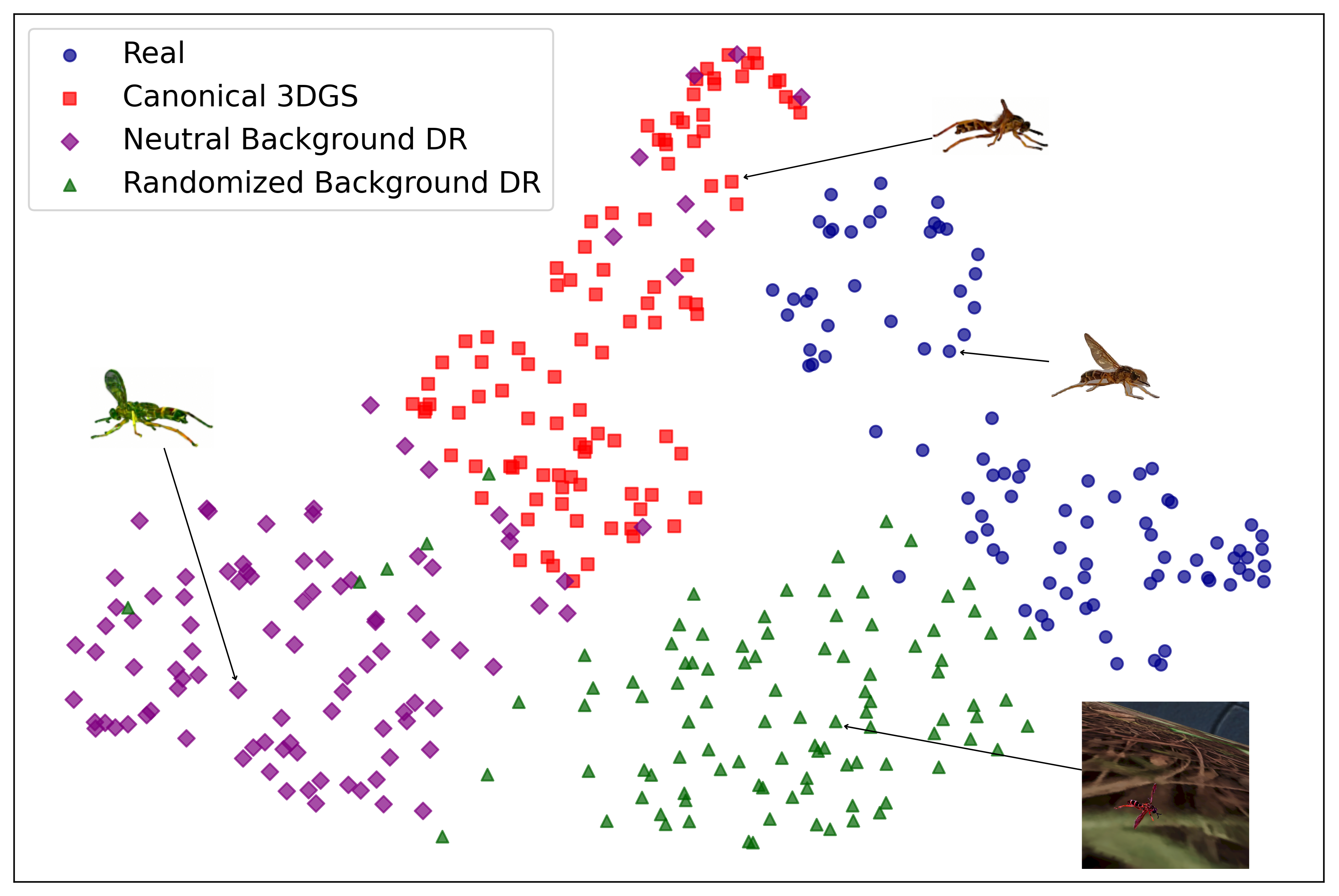}
        \caption{Fly specimen}
        \label{fig:umap_fly}
    \end{subfigure}
    \caption{2D UMAP feature-space projections of pre-trained ResNet-50 embeddings across four categories: Original video frames with backgrounds removed (dark blue circle), unperturbed canonical 3DGS renderings (red square), neutral-background DR renderings (purple diamond), and randomized-background DR renderings (green triangle). In both specimens, the unperturbed canonical models form distinct,  grouped clusters that exhibit an expected render gap relative to the real photographs. This baseline separation is eclipsed by the topological dispersion introduced by the photometric and procedural DR pipelines, increasing the dataset variance required for robust sim-to-real generalization.}
    \label{fig:umap_domain_coverage}
\end{figure}

As illustrated in the UMAP projections for both the moth (Fig.~\ref{fig:umap_moth}) and the fly (Fig.~\ref{fig:umap_fly}), the embeddings of the canonical 3DGS renderings form distinct, grouped clusters. While visually similar to the original captures, these clusters exhibit a spatial offset relative to the real-world images. This separation is expected and reflects the sim-to-real gap typical of synthesized media, arising from the smoothing of high-frequency textures by Spherical Harmonics and from underlying rasterization artifacts. Nevertheless, the density of these canonical clusters indicates that the baseline models remain morphologically consistent.

The scale of this initial domain shift, however, is negligible when compared to the dispersion introduced by our Domain Randomization pipelines. One can observe that the embeddings from the randomized 3DGS models expand the feature space domain, overcoming the textural and illumination constraints of the original captures. Ultimately, this broad feature-space coverage provides the varied training data necessary to encourage sim-to-real generalization.

To quantitatively corroborate our visual UMAP analysis, we computed the Silhouette Score~\cite{ROUSSEEUW198753} to assess the cluster separation between the real and synthetic datasets in the high-dimensional feature space. For a given embedding $i$, the silhouette score $s(i)$ is defined as:
\begin{equation}
    s(i) = \frac{b(i) - a(i)}{\max\{a(i), b(i)\}},
\end{equation}
where $a(i)$ represents the mean intra-cluster distance between point $i$ and all other points within its own category, and $b(i)$ denotes the mean nearest-cluster distance to all points in the closest neighboring category. The global metric $\bar{s}$ is the arithmetic mean of $s(i)$ across all $n$ points.

We evaluated $\bar{s}$ for both the moth and fly specimens under two distinct configurations: (1) segmented original real images versus the unperturbed canonical 3DGS models, and (2) segmented original real images versus the rendered models from the full DR pipeline. A score close to $0$ indicates highly overlapping or adjacent clusters, whereas a score close to one indicates greater spatial separation.

For the canonical baseline comparison, the evaluation yielded a low $\bar{s}$ of 0.1436 for the moth and 0.1532 for the fly. These low values confirm the narrow render gaps observed in the UMAP projections, reinforcing that the unperturbed 3DGS reconstructions act as reliable, morphologically proximate representations of the real specimens. In contrast, when evaluating the full DR configuration, $\bar{s}$ increased to 0.2235 for the moth and 0.1847 for the fly. This measurable increase confirms that the DR pipeline successfully expands the feature-space variance. Importantly, the conservative magnitude of this increase indicates a controlled expansion: while sufficient visual diversity is introduced to overcome textural biases, the underlying morphological identity of the specimens is preserved, ensuring the synthetic data remains semantically anchored to the real domain.
  
\subsection{Limitations}
While our Domain Randomization framework successfully expands the visual variance of 3DGS representations, it is subject to technical and physical constraints. 

First, an optical limitation arises from the use of macro-photography. Capturing small biological specimens with casual multi-view setups on conventional mobile devices results in shallow depth of field and defocus blur across video frames. Mitigating this traditionally requires hardware-intensive focus-stacking protocols, which penalize acquisition throughput and are prohibitive for dynamic captures~\cite{Berger2025}. Our canonical 3DGS reconstructions inherit these native blurring artifacts, which can degrade the baseline fidelity of the optimized radiance fields.

Second, the geometry in our framework remains rigid. Although we can continuously randomize appearance to overcome texture biases, we cannot currently animate articulated anatomical poses, such as varying insect wing dihedral angles or leg joints. 

Third, standard 3DGS features an entangled representation of materials and illumination. Because the explicit radiance field bakes in the original environmental lighting, the imported specimens do not dynamically respond to the rasterization engine's native physics-based lighting systems (e.g., Unity's directional lights or real-time shadows). Furthermore, within our photometric perturbation pipeline, correctly manipulating and rotating the illumination is mathematically straightforward for the zero- and first-order Spherical Harmonics (SH). However, accurately rotating higher-order SH coefficients to preserve complex, high-frequency view-dependent specularities under arbitrary spatial re-orientation remains computationally challenging. Consequently, our pipeline relies on intensity scaling for these higher orders or limits explicit rotations to lower-order bands.

Finally, from a software architecture perspective, the current implementation operates as a decoupled, two-stage pipeline. The stochastically perturbed 3DGS parameter tensors are precomputed in Python and exported as individual \texttt{.ply} files before being ingested into the Unity engine. Developing a unified architecture where tensor manipulation and rasterization occur simultaneously in GPU memory within the same engine environment is required to eliminate disk I/O bottlenecks and maximize dataset generation throughput.

\section{Conclusion}
We presented a meshless Domain Randomization framework that leverages the explicit, continuous parameter space of 3D Gaussian Splatting to generate synthetic data. By perturbing the unstructured radiance representation, modulating photometric properties, and overriding native albedos with procedural noise, our approach increases visual variance in the training data. By compositing these randomized fields in a custom rendering engine, we provide zero-cost 2D annotations for arbitrary camera poses. Our feature-space evaluations indicate that direct tensor manipulation broadens the synthetic embeddings' feature domain, overcoming the narrow texture and lighting constraints of the original captures while preserving the essential morphological identity of the specimens.

As future work, our immediate priority is to evaluate the proposed synthesized datasets by training state-of-the-art taxonomic object detectors (e.g., YOLO variants). We aim to assess their zero-shot sim-to-real transfer capabilities by identifying bottlenecks in detecting insects that exhibit strong natural camouflage against complex ecological backgrounds. While our current procedural DR expands the range of general visual variation, overcoming the textural biases evident in these ecological scenarios remains a critical challenge. 

Additionally, to overcome the optical constraints of casual multi-view captures without relying on restrictive laboratory setups, we plan to investigate generative image-to-image restoration models applied to the input frames. By removing defocus blur from the raw captures prior to the 3DGS optimization, we can refine the training views and ensure more reliable convergence of the radiance field. We also plan to integrate articulated 3DGS representations \cite{hu2024gauhuman} to enable anatomical pose randomization alongside our structural perturbations. Finally, consolidating the decoupled parameter-manipulation and rasterization stages into a unified engine architecture remains a key engineering objective for maximizing synthesis throughput.

\section*{Acknowledgment}
This study was financed in part by  CAPES - Finance Code 001, and CNPq (304027/2022-7, 403280/2025-7, 302259/2026-0). The authors also thank Daubian Santos for collecting the insect images. 

\section*{AI Assistance Declaration}
The authors declare the use of Grammarly Pro and Google Gemini to assist with language refinement, LaTeX formatting, and code refactoring. The authors assume full responsibility for the accuracy, originality, and final content of this paper.

\bibliographystyle{IEEEtran}
\bibliography{ref}

\end{document}